# Wetting-Layer-Assisted Synthesis of Inverted CdSe/PbSe Quantum Dots and their Photophysical and Photo-Electrical Properties

*Vladimir Sayevich, Whi Dong Kim, Zachary L. Robinson, Oleg V. Kozlov, Clément Livache, Namyoung Ahn, Heeyoung Jung, and Victor I. Klimov[*]*

Nanotechnology and Advanced Spectroscopy Team, C-PCS, Chemistry Division, Los Alamos National Laboratory, Los Alamos, New Mexico 87545, USA

E-mail: klimov@lanl.gov

**Abstract**

Heterostructured quantum dots (QDs) based on narrow-gap PbSe and wide-gap CdSe have been studied with an eye on their prospective applications in near-infrared (NIR) light sources, photodetectors, and solar cells. The most common structural motif is a spherical QD comprising a PbSe core enclosed into a CdSe shell. However, the potential barrier created by the CdSe shell complicates extraction of band-edge charge carriers from the QD. Therefore, conventional PbSe/CdSe QDs are not suitable for applications in practical photoconversion devices. Here we report inverted CdSe/PbSe core/shell QDs that overcome this drawback. In these structures, both photocarriers (electron and hole) exhibit a significant degree of shell localization and are therefore free to move within the QD solid and be extracted into an external circuit. To create such QDs, we employ a novel synthetic method in which a thin, atomically controlled wetting layer is used to homogenize the surface of the CdSe core and thus promote directionally uniform growth of the PbSe shell. Unlike noninverted QDs, inverted core/shell structures exhibit highly efficient photocarrier transport, making them excellent candidates for applications in practical photoconversion including photovoltaics, photodetection, and photochemistry.

# 1. Introduction

By combining different materials in a single nanocrystalline heterostructure, it is possible to realize properties not accessible with monocomponent systems.[1-3] In the case of colloidal hetero-nanocrystals, or quantum dots (QDs), a common structural motive is a core/shell or a core/multi-shell QD. Current methods to synthesize such structures are based on direct growth of one semiconductor on top of a core made of other material(s) or anion/cation exchange within a peripheral layer of a pre-synthesized core. The first of this method was originally implemented by overcoating a CdSe QD with a layer of wider-gap ZnS which resulted in core/shell structures with an increased emission efficiency, a direct result of improved passivation of surface dangling bonds.[4] Later, this method was successfully adopted for many other combinations of core and shell materials.[1-3] In addition to improving the emission efficiency, the core/shell approach allows manipulation of other QD functionalities, including the character of the band-edge transition (direct *versus* indirect),[5,6] the exciton-exciton interaction energy,[7] the Auger recombination rate,[8-10] and the emission intermittency of a single QD emitter.[11,12]

The ion exchange method is usually applied to prepare PbSe(S)/CdSe(S) core/shell QDs.[13-16] These heterostructures have been extensively studied as high-efficiency infrared (IR) light emitters.[17-19] In principle, they are also attractive for application in photoelectric devices, including advanced systems exploiting carrier multiplication[20] (CM) and up-conversion[21,22] concepts. However, the realization of such devices with conventional core/shell PbSe/CdSe QDs is not straightforward as photogenerated carriers are confined to the PbSe core, which complicates their extraction from the QD due to a large potential barrier created by the CdSe shell. This barrier is especially high (~1.4 eV) for photoholes,[16,20] which makes them not readily accessible electrically.

The problem of charge extraction can, in principle, be resolved by inverting the nanostructure, that is, by preparing CdSe(core)/PbSe(shell) QDs. In such inverted structures, wavefunctions of both an electron and a hole are expected to extend to the QD surface which should simplify their extraction from the QD. The inverted CdSe/PbSe QDs have been pursued previously, however, without considerable success. Application of traditional cation exchange usually results in either complete replacement of $Cd^{2+}$ for $Pb^{2+}$ ions in the entire QD volume[23] or partial asymmetric cation exchange leading to segmented heterostructures such as Janus particles.[24] The difficulty to obtain spherically symmetric core/shell QDs with the inverted CdSe(S)/PbSe(S) geometry is at least partially related to a considerable difference in reactivities of different crystal facets of a CdSe core compounded by nonuniformities in the ligand coverage.

Here we report a synthetic method – wetting-layer-assisted deposition (WLAD) – which allows us to fabricate symmetric hetero-QDs with an inverted CdSe(core)/PbSe(shell) geometry. The key element of the WLAD technique is a thin atomically controlled wetting layer (WL) of PbSe prepared on top of a CdSe core using a self-limiting colloidal atomic layer deposition (c-ALD). This approach allows us to eliminate surface nonuniformities arising from a facet-dependent crystal structure and/or a nonuniform composition/density of a molecular ligand layer. Due to the introduction of the WL, we obtain a uniform, highly reactive core surface, which facilitates directionally symmetric growth of the PbSe shell of a controlled thickness. We show that the WLAD technique can be successfully applied to prepare inverted core/structures using starting CdSe cores with both a wurtzite (WZ) and zinc-blende (ZB) crystal structure. We study the fabricated inverted QDs using time-resolved photoluminescence (PL) and transient photocurrent (TPC) spectroscopies. The PL spectra exhibit a visible and an IR emission features, as previously observed for conventional (noninverted) PbSe/CdSe core/shell QDs. The visible band shows fast decay which reflects rapid relaxation of a 'hot' CdSe-core-localized exciton into lower-energy states with the PbSe shell. The TPC measurements reveal efficient hole-dominated charge transport which is similar to that of standard PbSe(S) QD solids. Thus, the developed inverted core/shell CdSe/PbSe QDs combine optical characteristics similar to those of conventional PbSe/CdSe QDs with good photocarrier transport properties derived from a favorable distribution of carrier wavefunctions that exhibit a considerable degree of shell localization. These properties make inverted CdSe/PbSe QDs of considerable utility in photoelectric devices in which absorbed light is converted into electric current. These structures may also be useful in solar photochemistry, as their near-infrared (NIR) bandgap is beneficial for harvesting sunlight, while the ease of extracting both electron and hole should simplify the use of photogenerated carriers in chemical reactions.

## 2. Quantum Dot Synthesis and Structural Characterization

The WLAD approach comprises two steps: (1) polar-phase-based deposition of a WL of a shell material (PbSe) onto pre-synthesized CdSe cores, followed by (2) the growth of a uniform PbSe shell in a non-polar phase at elevated temperature performed by adding controllable amounts of molecular precursors (Figure 1; for the details of the synthetic procedure see the Experimental Section and Scheme S1, Supporting Information).

Briefly, first, we use established non-polar-phase syntheses to prepare highly monodisperse CdSe cores with either WZ[25] or ZB[26] lattices (Figure 1a). The fabricated nanocrystals are transferred into a polar solvent (*N*-methylformamide, MFA), which is accompanied by the replacement of original organic ligands (mostly aliphatic alkylcarboxylates and phosphates) with $Se^{2-}$ species (Figure 1b). Negatively charged $Se^{2-}$ ionic ligands feature high binding affinity to electrophilic moieties on a CdSe QD surface, which helps remove all original organic molecules independent on the strength of their bonding to the nanocrystal surface and/or the attachment mode.[27] In particular, as indicated by measurements using a Fourier transform infrared (FTIR) spectroscopy (Figure S1, Supporting Information), FTIR features due to the C-H starching modes are pronounced in the original samples. However, they completely disappear from the FTIR spectra following organic-for-inorganic ligand exchange.

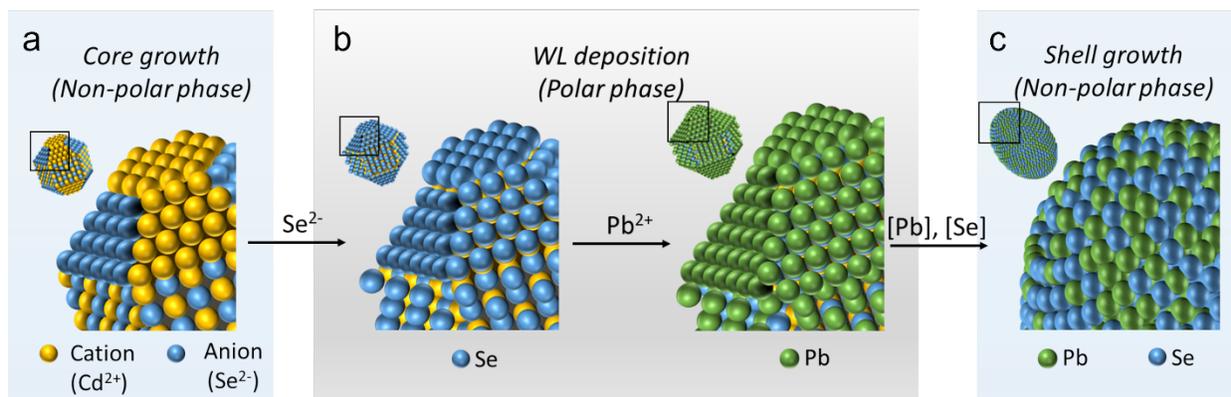

Figure 1. Schematic representation of the synthesis of inverted CdSe/PbSe core/shell QDs using the wetting-layer assisted deposition (WLAD) method. (a) The first reaction step is the standard non-polar-phase synthesis of CdSe cores with ZB or WZ crystal structures. (b) This is followed by transfer of the synthesized cores into a polar solvent, accompanied by the replacement of the original organic ligands with ionic $Se^{2-}$ species. The $Se^{2-}$ terminated QDs are then reacted with $Pb^{2+}$ cations, resulting in the formation of a highly unform PbSe WL. (c) In the third step, the WL-enclosed QDs (CdSe+WL) are transferred into a nonpolar phase and reacted with controlled amounts of Pb and Se molecular precursors in the presence of oleate co-ligands to prepare a PbSe shell of the desired thickness.

After the $Se^{-2}$-terminated CdSe cores are transferred into a polar solvent, they are reacted with $Pb^{2+}$ using c-ALD, which leads to the formation of a Se-Pb WL (Figure 1b). This layer is highly uniform and comprises exactly one semiconductor monolayer (ML) of a targeted shell material. Due to their small sizes, $Se^{2-}$ and $Pb^{2+}$ ions of the WL can easily adapt an irregular surface morphology of a faceted colloidal nanocrystal which helps homogenize the particle surface and thereby facilitate follow-up isotropic growth of the PbSe shell. The particles produced at this stage are further referred to as (CdSe+WL) functionalized cores.

To continue shell growth, we add lead (II) oleate to the $Pb^{2+}$-terminated functionalized cores and transfer them back into a nonpolar phase (Figure 1c). 'Soft' oleate anions feature good affinity to 'soft' surface $Pb^{2+}$ cations, which is combined with their high lability at elevated temperatures.[28] The former property facilitates transfer of the particles into a nonpolar medium, while the latter simplifies simultaneous nucleation of PbSe at multiple sites over the entire nanocrystal surface which is essential for promoting uniform shell growth.[29,30] Following transfer into a nonpolar solvent, we react the (CdSe+WL) structures with controlled quantities of precursors of Se (trioctylphosphine selenide, TOP-Se) and Pb (lead oleate, Pb-OL or lead chloride oleylamine, $PbCl_2 \cdot OLA$) to achieve a desired shell thickness (Figure 1c).[31] The amounts of Pb and Se stock solutions were calculated from the core and shell volumes using bulk lattice parameters of PbSe and CdSe.

To grow the shell we prepare two syringes, one of which is filled with thoroughly washed functionalized (CdSe+WL) cores dissolved in octadecene (ODE) and another with the solution of the Se-precursor. Then, we inject the contents of both syringes into a mixture of the Pb precursor, ODE, and oleic acid (OA) heated to temperature $T$ = 130–140 °C. Afterwards, the temperature of the reaction mixture is raised to 150 °C for the ZB cores or to 160 °C for the WZ cores. The reaction is allowed to proceed for 0.5 – 4 min and then quenched by fast cooling. The longer reaction time leads to a thicker PbSe shell. This method allows us to grow the PbSe shell with a thickness ($H$) of up to ~1.2 nm, which corresponds to ~4 semiconductor MLs. Thicker PbSe shells were obtained by additional injections of Pb- and Se-precursors into the solution of the CdSe/PbSe QDs at 140 °C (Scheme S1, Supporting Information).

Transmission electron microscopy (TEM) measurements of the initial ZB CdSe cores and the final inverted core/shell CdSe/PbSe QDs are shown in Figure 2a. The starting CdSe cores are characterized by a mean radius (*r*) of 2 nm and a standard size deviation (δ*r*) of 0.14 nm, which corresponds to δ*r*/*r* of ~7% (Figure 2b, blue). Following the deposition of the PbSe shell, the particle radius increases to *R* = 2.9 nm, and the size dispersion (δ*R*/*R*) broadens to ~14% (Figure 2b, red). Using the core and core/shell QD sizes, we find that the average shell thickness is *H* = 0.9 nm or ~3 PbSe MLs. Based on the analysis of size polydispersities (δ*r* and δ*R*), the standard deviation of shell thickness (δ*H*) is ~0.38 nm or ~1 ML.

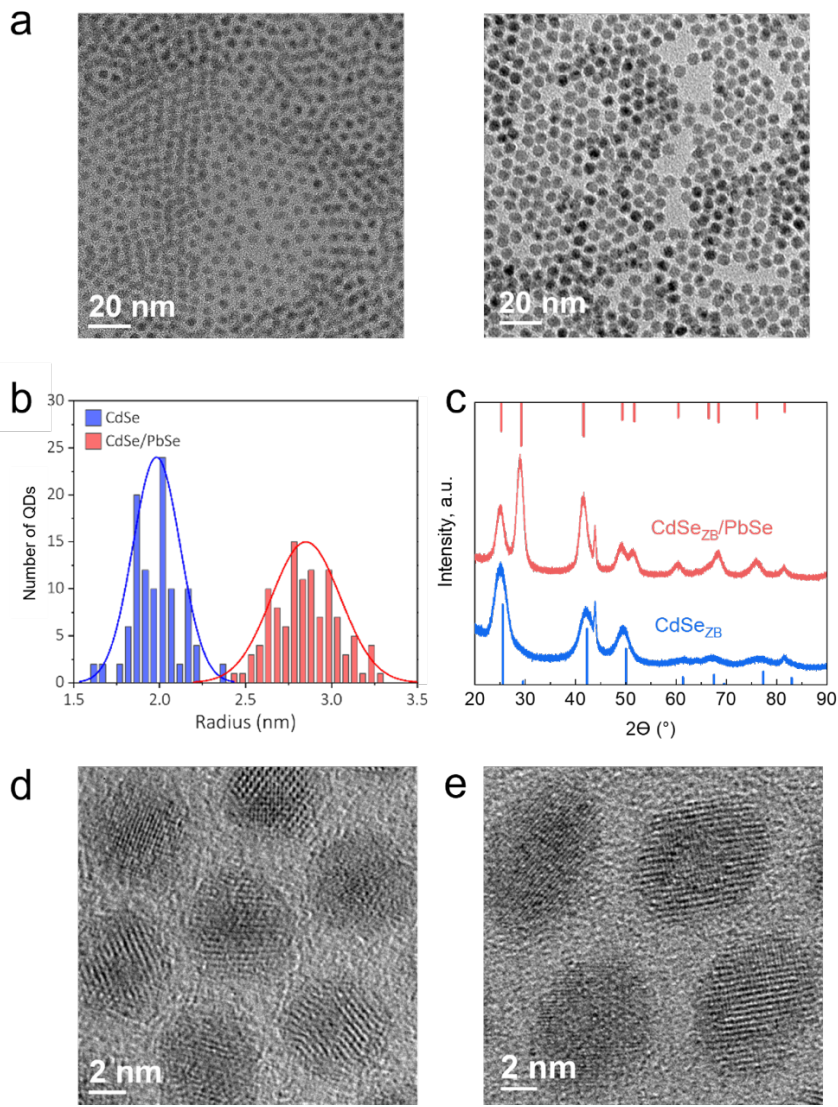

Figure 2. (a) TEM images of the starting ZB CdSe cores (left) and the final core/shell CdSe/PbSe QDs (right). (b) Size distributions of the ZB CdSe cores (blue bars) and final core-shell QDs (red bars). Based

on a Gaussian fit (color-matched lines), mean radii of the cores and the core/shell structures are 2.0 nm and 2.9 nm, respectively. The corresponding standard deviations are 0.14 nm (7%) and 0.4 nm (14%), respectively. (c) XRD patterns of the starting ZB CdSe cores (blue) and core/shell CdSe/PbSe structures (red). Vertical bars show XRD peaks of bulk ZB CdSe (bottom, blue) and rock salt PbSe (top, red). (d,e) High-resolution TEM images of inverted CdSe/PbSe QDs prepared using ZB (d) and WZ (e) CdSe cores. Their mean total radii are 2.8 nm and 3.5 nm, respectively. From the combined analysis of the ICP-EOS and TEM data (see text for details), the CdSe core radii for these QDs are 1.1 nm and 1.45 nm, respectively. Based on these values, the PbSe shell thicknesses are, respectively, 1.7 nm and 2.05 nm or 5.6 and 6.7 MLs.

The analysis of the crystal structure using X-ray diffraction (XRD) indicates that the cores exhibit the expected features of a ZB CdSe crystal lattice (Figure 2c, blue trace). The XRD measurements of the core/shell QDs show the predominance of peaks characteristic of the PbSe rock-salt lattice (Figure 2c, red trace). This is a consequence of a larger volume of the PbSe shell compared to that of the CdSe core in the final core/shell structures.

In Figure S2 (Supporting Information), we show examples of TEM images and XRD patterns of the CdSe cores ($r$ = 1.9 nm) and corresponding CdSe/PbSe structures ($R$ = 3.5 nm) for samples grown using WZ cores. Both the core-only and core/shell QDs exhibit good size uniformity, comparable to that of the ZB-core samples. The XRD measurements show a transition from the WZ pattern for the core-only sample to the rock-salt pattern for the core/shell QDs. As indicated earlier, this occurs due to a larger volume of the PbSe shell compared to that of the CdSe core.

Examination of the high-resolution TEM images in Figure 2d,e indicates that the shape of a core/shell QD closely replicates the shape of the original cores. In particular, the ZB CdSe cores exhibit a nearly spherical shape (Figure 2a) while the WZ cores are slightly elongated (Figures S2,S3, Supporting Information), as observed previously.[32] These shapes are preserved following the formation of the CdSe/PbSe hetero-structures (Figure 2d,e). This points towards highly uniform deposition of the shell material whose rate does not vary significantly between different nanocrystal facets. This is a direct consequence of our WL approach, which helps homogenize the core surface and, as a result, leads to good uniformity of the shell thickness.

Interesting information about the details of shell growth is provided by the comparison of TEM-based size measurements and the compositional analysis using inductively coupled plasma optical emission spectroscopy (ICP-OES) (Table 1). To discuss this comparison, we consider the case of the structures grown using WZ CdSe cores. As indicated earlier, the WZ core and resulting core/shell QDs have elongated shape. We define their size in terms of the effective (volume-equivalent) radius determined as $R = (a^2b)^{1/3}$, where $a$ and $b$ are the semi-major and semi-minor axes, respectively. For the series of samples used in the ICP-OES studies, the CdSe core radius was 1.9 nm. After the preparation of a PbSe WL, the particle radius increased to ~2.2 nm and then, to ~2.9 nm following a continuing shell growth in a nonpolar medium using a 4-min reaction with Pb (II) oleate- and TOP-Se. The addition injection of the PbSe shell precursors produced the particles with $R = 3.5$ nm which completed the sample series.

Based on the ICP-OES measurements (Table 1), we can obtain the ratio of the numbers of the Pb ($N_{Pb}$) and Cd ($N_{Cd}$) ions ($\beta = N_{Pb}/N_{Cd}$) and then, use this quantity to calculate the core radius from $r = R[1 + \beta v_{PbSe}/(2v_{CdSe})]^{-1/3}$. Here $v_{CdSe}$ and $v_{PbSe}$ are the unit cell volumes of bulk WZ CdSe and rock salt PbSe, respectively ($v_{CdSe} = 0.112$ nm$^3$, $v_{PbSe} = 0.236$ nm$^3$). A factor of 2 is added in front of $v_{CdSe}$ because the rock-salt PbSe unit cell contains twice as many cations as the WZ CdSe unit cell. Using this expression, we obtain $r$ of 1.9 to 2.0 nm for the PbSe shell growth time ($t_{sh}$) of up to 3 min ($H$ to about 0.7 nm), suggesting that the core retains its original size. For $t_{sh} = 4$ min, the CdSe core radius determined from ICP-OES measurements drops to 1.7 nm ($R = 2.9$ nm) and further to 1.45 nm following the additional injection of shell precursors ($R = 3.5$ nm). These observations suggest that the prolonged shell growth is accompanied by a partial Cd-for-Pb cation exchange, resulting in a gradual decrease in the CdSe core size.

Table 1. TEM and ICP-OES measurements of a series of inverted core/shell CdSe/PbSe QD samples with different PbSe shell thicknesses. These samples were prepared using WZ CdSe cores with $r$ = 1.9 nm.

| Shell growth time (min) | $\beta = N_{Pb}/N_{Cd}$ (ICP-OES) | QD total radius (nm) (TEM) | CdSe core radius* (nm) | PbSe shell thickness** (nm) |
|---|---|---|---|---|
| 0 | | | 1.9 | 0 |
| 0.5 | 0.71±0.05 | 2.3±0.3 | 1.9 | 0.4 |
| 1 | 0.85±0.06 | 2.5±0.2 | 2.0 | 0.5 |
| 2 | 1.22±0.09 | 2.6±0.2 | 2.0 | 0.6 |
| 3 | 1.52±0.11 | 2.7±0.2 | 2.0 | 0.7 |
| 4 | 4.1±0.3 | 2.9±0.2 | 1.7 | 1.2 |
| Additional injection | 12.3±0.9 | 3.5±0.3 | 1.45 | 2.05 |

*CdSe core radius ($r$) was calculated using $r = R[1 + \beta v_{PbSe}/(2v_{CdSe})]^{-1/3}$, where $v_{PbSe}$ and $v_{CdSe}$ are the unit cell volumes of bulk WZ CdSe and rock-salt PbSe, respectively.

**PbSe shell thickness ($H$) was obtained from $H = R - r$.

## Optical Spectroscopy Studies

For spectroscopic measurements, the QDs were purified using acetone as an aprotic solvent in air-free environment to reduce a detrimental effect of oxygen on PbSe-shelled QDs, redissolved in anhydrous toluene, and loaded into an optical cell with a thickness of 1 mm. Steady-state PL spectra were measured using low-intensity continuous wave (*cw*) excitation at 532 nm. Time-resolved PL measurements were performed using 100-fs pump pulses with a photon energy of 1.5 eV (NIR PL feature) or 3 eV (visible PL feature). To avoid complications caused by multiexciton effects, we used a low per-pump fluence, corresponding to a sub-single-exciton average QD occupancy. The PL signal was time-resolved using a superconducting nanowire single-photon detector (SNSPD) with a temporal resolution of 60 ps.

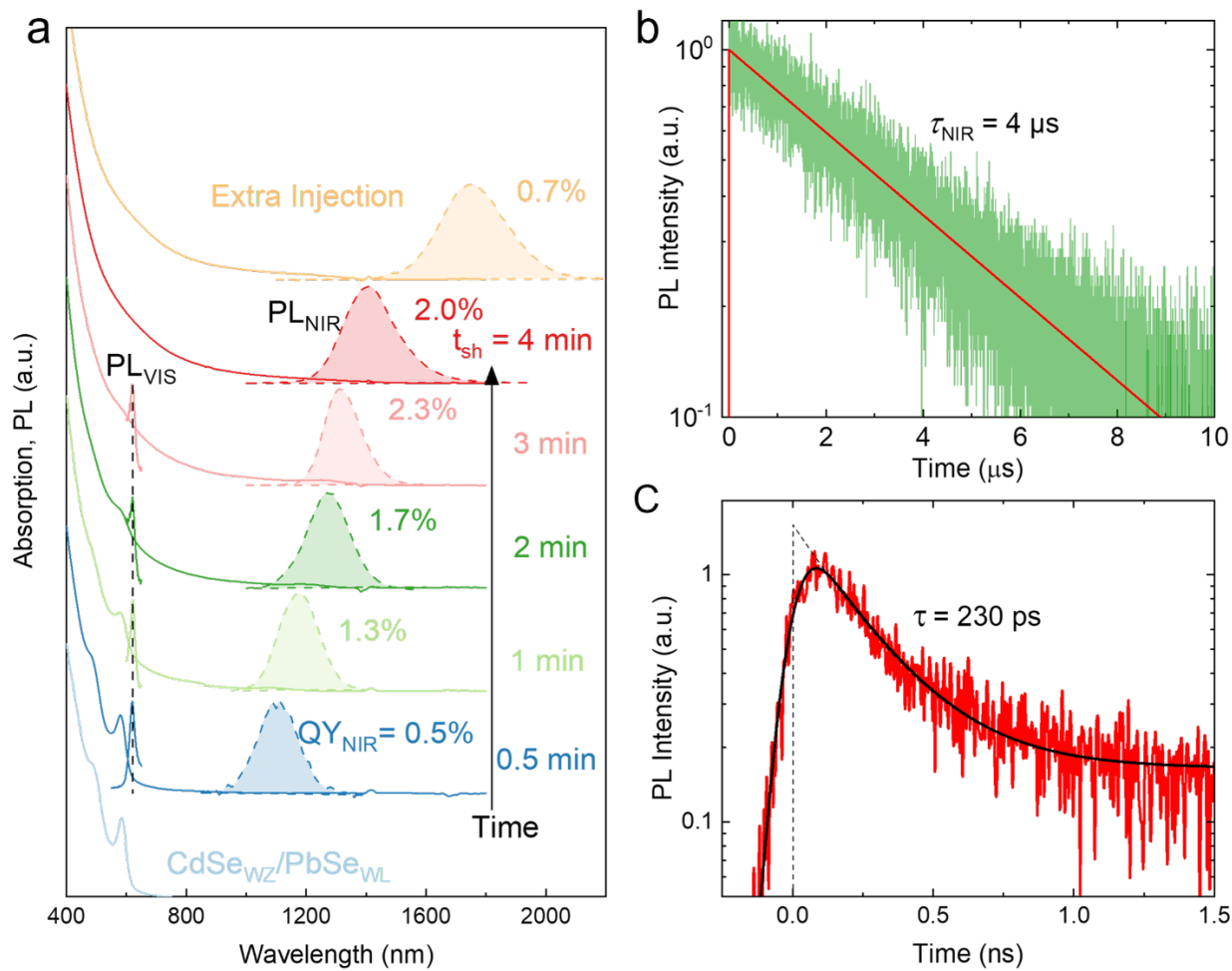

Figure 3. (a) Evolution of linear absorption and PL spectra of inverted core/shell CdSe/PbSe QDs with increasing shell thickness. These samples were prepared using WZ CdSe cores with $r$ = 1.9 nm. The thickness of the PbSe shell ($H$) changes from 0.4 nm (shell growth time $t_{sh}$ = 0.5 min) to 1.24 nm ($t_{sh}$ = 4 min) and then to 2.05 nm (after extra injection of Pb and Se precursors) (b) PL dynamics of CdSe/PbSe QDs with $H$ = 0.5 nm ($t_{sh}$ = 1 min) measured at the peak (~1200 nm) of the NIR band (green). Red line is a single exponential fit with a time constant of $\tau_{NIR}$ = 4 μs. (c) The visible PL dynamics measured at 619 nm (red line) is fitted to a double exponential decay (dashed black line) convolved with the instrument response function of the SNSPD apparatus, yielding the trace shown as the solid black line. Based on the fit, the initial (fast) relaxation time constant is 230 ps.

Figure 3a and Figure S3 (Supporting Information) show optical absorption and PL spectra of core-only and core/shell CdSe/PbSe QDs from the same series of samples that were used in the ICP-OES studies (Table 1). The original CdSe core particles ($r$ = 1.9 nm) exhibit a sharp band-edge absorption peak at 576 nm and a fairly narrow PL band at 588 nm with a full width at half maximum (FWHM, $\Gamma_{PL}$) of 25 nm or 89 meV (Figure S3a, Supporting Information). After application of a WL, the particles become non-emissive but still exhibit a sharp band-edge absorption peak slightly shifted to longer wavelengths (the peak position is 584 nm; Figure 3a and Figure S3a, Supporting Information). This indicates the increase in the carrier effective localization volume due to expansion of carrier wavefunctions into the WL.

After particle transfer into the nonpolar phase followed by the 0.5-min shell-growth reaction (resulting in $H$ of ~0.4 nm), the QD sample exhibits dual-band PL consisting of a weak visible peak at 619 nm and a stronger NIR feature at $\lambda_{NIR}$ = 1100 nm with a quantum yield ($QY_{NIR}$) of ~0.5% (Figure 3a). As indicated by our modeling of electronic states (discussed in the next section), these two features are due to a common band-edge electron state and two different hole states. The NIR band corresponds to a band-edge hole with preferential shell localization, while the visible band is due to a higher energy hole state which is primarily core localized. The absorption spectrum also shows signatures of core and shell components of the hetero-QDs. Specifically, it exhibits a structureless low-energy tail arising from the PbSe-shell-based states and the CdSe-core related peak at 584 nm.

For longer shell deposition times ($t_{sh}$ = 1 to 3 min; the corresponding $H$ is 0.5 to 0.7 nm), the CdSe core-related absorption feature becomes broader and less pronounced (Figure 3a). This is accompanied by a progressive increase of the intensity of the low-energy absorption tail arising from the PbSe shell. At the same time, the NIR PL band shifts towards lower energies (reaching 1400 nm for $t_{sh}$ = 3 min) and increases in intensity ($QY_{NIR}$ = 2.3% for $t_{sh}$ = 3 min). Further increase of the shell thickness to 1.2 nm ($t_{sh}$ = 4 min) and then to 2.05 nm (additional injection of shell precursors) leads to a continued redshift of the NIR PL band, but its intensity becomes weaker. For the thickest-shell sample ($H$ = 2.05 nm, $R$ = 3.5 nm), $\lambda_{NIR}$ = 1750 nm and $QY_{NIR}$ = 0.7%.

The change in the trend describing the dependence of $QY_{NIR}$ on $H$ correlates with the onset of the CdSe core etching due to the Cd-for-Pb cation exchange inferred from the ICP-OES studies (Table 1). This suggest that this process is accompanied by the formation of interfacial defects, which act as nonradiative decay centers. This also explains the complete quenching of visible PL at shell thicknesses greater than ~1 nm (Figure 3a).

Time-resolved PL measurements reveal a dramatic difference in relaxation time scales of NIR and visible emission bands (Figure 3b,c). The NIR PL feature exhibits slow microsecond decay with a relaxation time constant of $\tau_{NIR}$ = 4 μs (Figure 3b), similar to that of traditional NIR-emitting PbSe QDs and noninverted PbSe/CdSe core/shell structures.[33,34] The visible band has much shorter decay. Using the trace obtained by the deconvolution of the measured PL dynamics with an instrument response function of our SNSPD detector, the visible PL relaxation time is $\tau_{VIS}$ = 230 ps (Figure 3c). This fast relaxation indicates the short-lived nature of at least one of the states (electron or hole) responsible for the visible emission. As mentioned earlier, this short-lived state likely corresponds to a higher-energy 'hot' hole localized mainly in the CdSe core, implying that $\tau_{VIS}$ is determined by the relaxation of the 'hot' hole to lower-energy states with preferential localization in the PbSe shell.

Interestingly, although the core-to-shell hole relaxation in our inverted core/shell structures is much faster compared to radiative decay, it is considerably slower than the intra-band relaxation of 'hot' carriers in standard core-only QDs. Indeed, typical intra-band relaxation time constants in CdSe or PbSe QDs range from hundreds of femtoseconds to a few picoseconds.[37,38] Since these time scales are orders of magnitude shorter than those of radiative decay, core-only QDs do not exhibit any discernable emission from 'hot' carriers in excited states. On the other hand, our inverted core/shell QDs show a readily detectable 'hot' PL which is a direct consequence of the rather slow hole transfer from core- to shell-based states. We discuss mechanisms responsible for the slowing down of 'hot' hole relaxation in the inverted CdSe/PbSe core/shell QDs in the next section, where we perform a theoretical analysis of their electronic states.

## 4. Modeling of Electronic States

In our modeling, we use a mesh-based method for solving the radial Schrodinger equation in the single-band effective-mass approximation (Figure 4).[39] We consider heterostructures with a WZ CdSe core and use bulk semiconductor bandgaps ($E_g$) of 1.74 eV (CdSe) and 0.28 eV (PbSe), and carrier effective masses of $0.13m_0$ (electrons) and $0.45m_0$ (holes) for CdSe, and $0.084m_0$ (electrons) and $0.07m_0$, (holes) for PbSe. Coulomb interactions were included using dielectric constants of 6.3 for CdSe, 22.9 for PbSe, and 1 for the external medium. We further assume that the conduction-band energy offset ($\Delta E_{CB}$) at the CdSe/PbSe is zero,[40] implying that the valence-band energy offset ($\Delta E_{VB}$) is 1.46 eV ($\Delta E_{VB} = E_{g,CdSe} - E_{g,PbSe}$).

In Figure 4a, we display an energy-level diagram of a core/shell structure with $r$ = 1.9 nm and $R$ = 2.9 nm, which corresponds to shell thickness $H$ = 1 nm (approximately 3 PbSe MLs). The QD states are labeled as '$nl$'. Here $l$ is the orbital momentum of the envelope wavefunction denoted as S, P, and D for states with $l$ = 0, 1, and 2, respectively, and $n$ is the state number in a series of states with the same $l$, which varies from 1 to infinity.

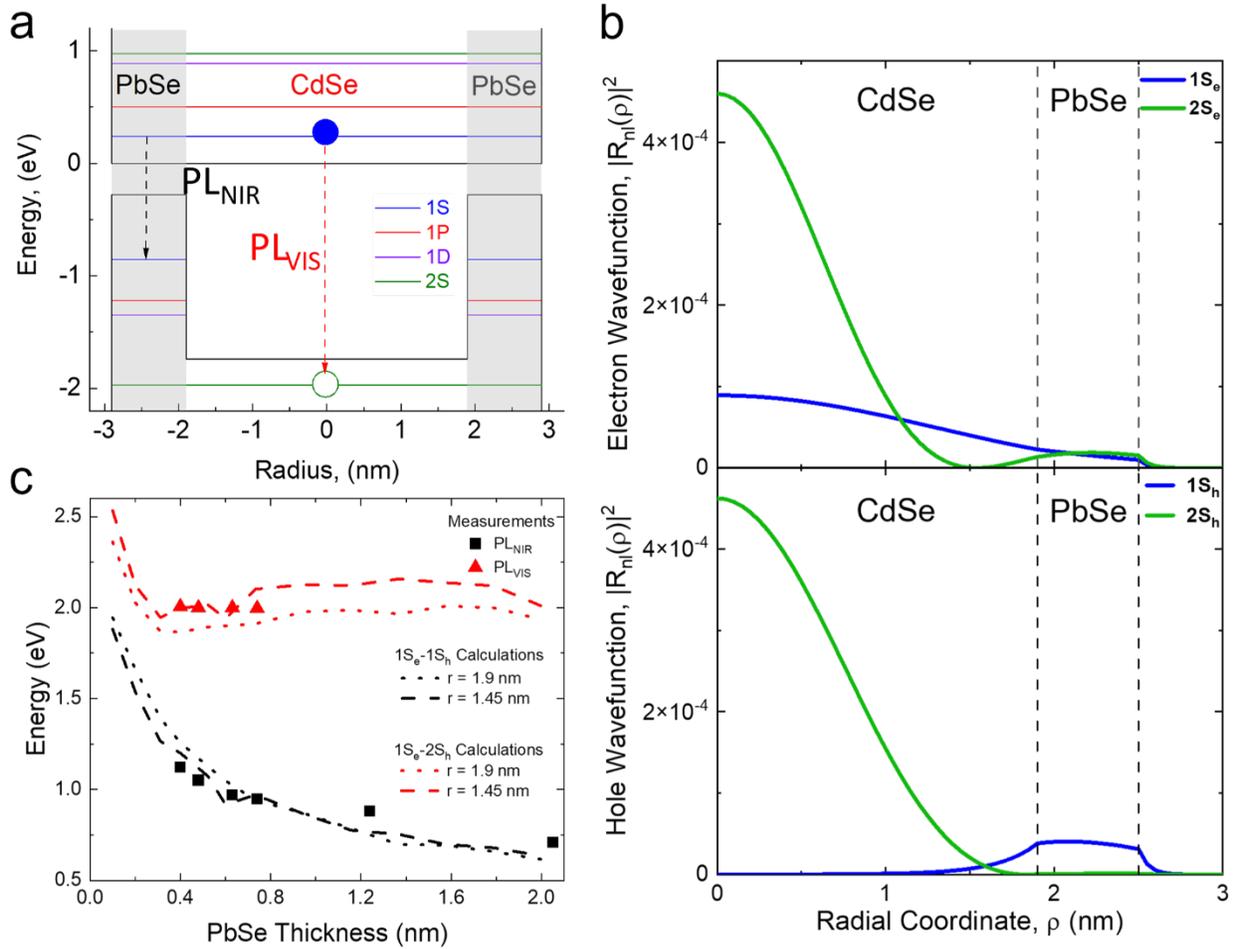

Figure 4. (a) Energy-level diagram of CdSe/PbSe QDs with $r$ = 1.9 nm and $H$ = 1 nm. The NIR emission band ($PL_{NIR}$, dashed black arrow) is due to the delocalized $1S_e$ electron recombining with the shell-localized $1S_h$ hole, while the visible emission band ($PL_{VIS}$, dashed red) is due to the delocalized $1S_e$ electron recombining with the core-localized $2S_h$ holes. (b) Calculated radial electron and hole wavefunctions for CdSe/PbSe QDs with $r$ = 1.9 nm and $H$ = 1 nm. The $1S_h$ and $2S_h$ states are localized preferentially within the shell and the core regions of the QD, respectively. (c) Calculated energies of the $1S_e$–$1S_h$, $1S_e$–$2S_h$ transitions (black and red lines, respectively) compared to the spectral energies of the NIR (black squares) and visible (red triangles) PL bands observed experimentally (Figure 3a). The core radii used in the calculations are $r$ = 1.9 and 1.45 nm. These correspond to the maximum and minimum CdSe radius values found in Table 1. The shrinking CdSe radius is due to cation exchange during the growth of thicker PbSe shells.

Figure 4b shows radial distributions of electronic wavefunctions for two S-type electron and hole states ($1S_e$, $2S_e$, $1S_h$ and $2S_h$) of these hetero-QDs (wavefunction calculations for lower-energy states of P- and D-type can be found in Figure S4, Supporting Information). Based on the calculations, all electron states are delocalized over the entire volume of a core/shell particle. In contrast, hole states show a pronounced difference in localization depending on their energy. In particular, the lowest energy $1S_h$ state is almost entirely shell localized and exhibits just weak leakage into the core region (Figure 4b, lower panel, blue line). Conversely, the higher-energy $2S_h$ state, located below the CdSe valence band-edge, is almost fully confined to the CdSe core (Figure 4b, lower panel, green line). As a result, the core- and shell-based hole states exhibit very weak wavefunction overlap. A similar consideration holds for the other two shell-localized hole states ($1P_h$ and $1D_h$), which also exhibit very weak overlap with the $2S_h$ wavefunction (Figure S4, Supporting Information). Together with the sizable energy gap between the $2S_h$ state and the nearest shell-based $1D_h$ level (Figure 4a), this impedes hole relaxation from the core to the shell and explains the emergence of narrow-band 'hot' PL arising from the $1S_e$–$2S_h$ transition.

The conducted calculations accurately describe the shell-thickness dependence of spectral energies of the observed NIR and visible PL bands ($E_{NIR}$ and $E_{VIS}$, respectively). In Figure 4c, we compare the measurements of $E_{NIR}$ and $E_{vis}$ (Figure 3a and Figure S3, Supporting Information) with modeling conducted for the inverted core/shell QDs with $r = 1.9$ nm and $r = 1.45$ nm, and a varied shell thickness. The two radii used in the calculations are the maximum and minimum values found in Table 1. As discussed earlier, the reduction of the CdSe core radius observed for thicker shell samples is due to cation exchange during the latter stages of PbSe shell growth.

We observed a good overall agreement between our calculations and measurements for both the NIR and visible bands (Figure 4c). This, in particular, indicates that the visible PL is indeed due to the $1S_e$–$2S_h$ transition, as suggested earlier. We would like to point out that the inclusion of the Coulomb interaction was necessary to obtain an accurate description of the experimental observations (Figure S5, Supporting Information). Otherwise, the calculated transition energies were higher than the observed PL energies. Since the electron-hole interaction is attractive, it reduces the energy of the interband transitions, which leads to the improved agreement with the measurements. The Coulomb correction is ~400 meV for a thin PbSe shell of 1 ML thickness and decreases to ~170 meV for a 7 ML shell (in both cases $r$ = 1.9 nm).

## 5. Transient Photocurrent Measurements

The regime of slow carrier cooling is a useful property of the inverted core/shell CdSe/PbSe QDs which is sought for various advanced photoconversion schemes including up-conversion of IR light,[22,41,42] 'hot-carrier' solar cells,[43] and CM.[44-46] The latter process can benefit both photovoltaics,[44,45,47] and photochemistry.[48]

The regime of slow carrier cooling was previously realized with traditional (noninverted) PbSe/CdSe core/shell structures[20] and PbS/CdS Janus particles.[49] In both type of the reported structures, it was used to enable high-efficiency CM wherein the kinetic energy of 'hot' slow-relaxing holes confined to the CdSe(S) region of the heterostructure was converted into an additional exciton. Even greater enhancement of the CM yield was recently achieved using Mn-doped noninverted PbSe/CdSe wherein the CM effect was driven not by direct but spin-exchange Coulomb interactions enabled by magnetic impurities.[50]

While being potentially useful in advanced photoconversion, traditional PbSe/CdSe QDs are not well suited for application in practical photoelectric devices. As discussed earlier, the primary problem is the difficulty to extract a band-edge hole which is confined to the PbSe core and thus is separated from the external medium by a large potential barrier created by the CdSe shell. This problem can be resolved with the inverted core/shell structures realized in the present study. As indicated by our modeling, in CdSe/PbSe QDs wavefunctions of 1S electrons and holes have a large amplitude at the QD surface (Figure 4b). Therefore, both types of carriers can readily communicate with the environment and, if necessary, be extracted from the QD and transported in the QD solid or the external circuit.

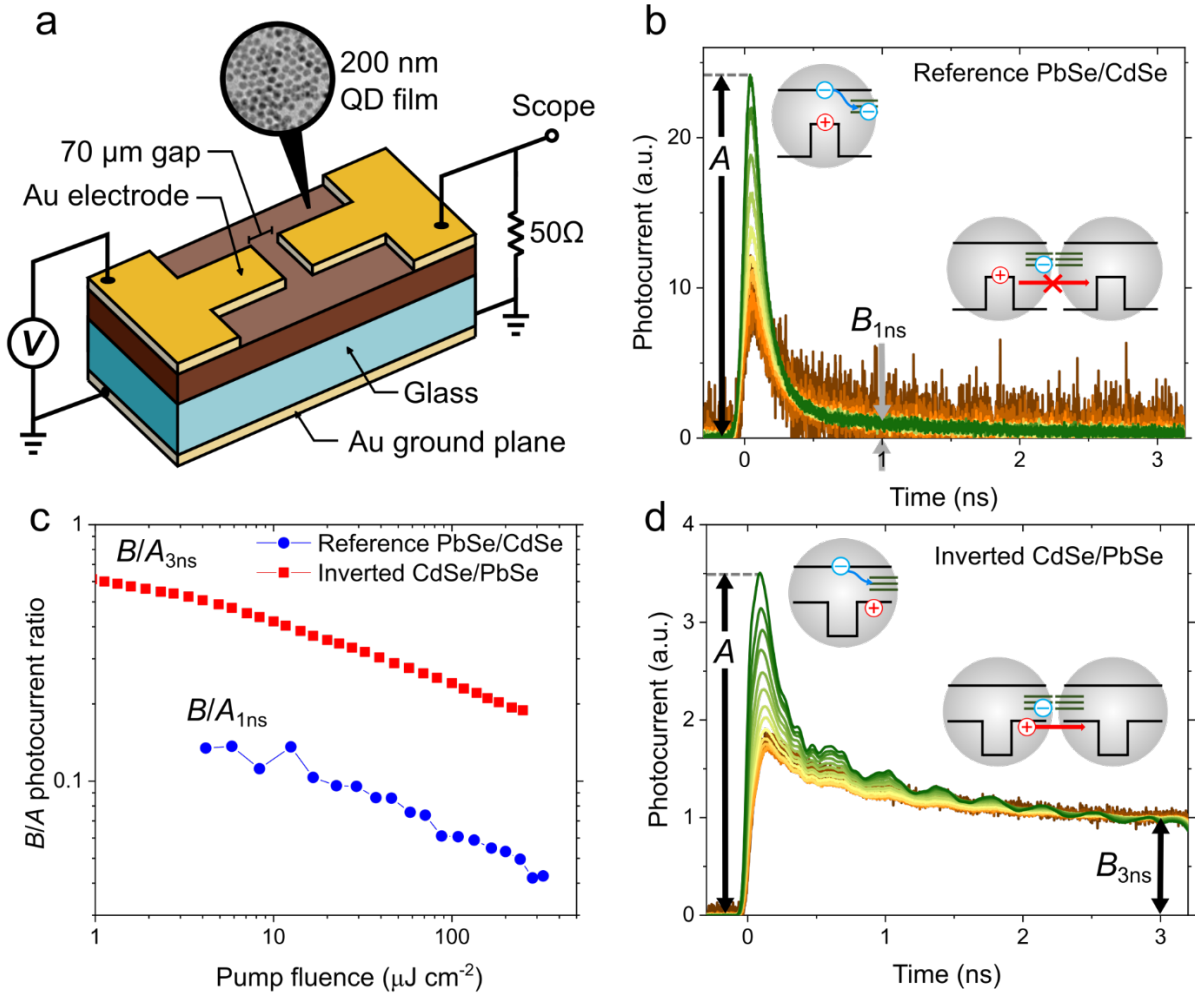

Figure 5. (a) A schematic depiction of a photoconductive Auston switch with a film of QDs as the photosensitive element. This device comprises a 200 nm thick QD film deposited onto a glass substrate with top gold electrodes that form a 50 Ohm transmission line together with a gold back plane. Photocurrent is generated in the 70 μm gap by short 110 fs laser pulses (1.2 eV photon energy) and monitored with a 20 GHz sampling oscilloscope. (b) Pump-power dependent tail-normalized TPC traces for EDT-treated reference (noninverted) PbSe/CdSe QDs (the total radius and shell thickness are ~4 nm and ~2.5 nm, respectively). The average per-dot excitonic occupancy $\langle N \rangle$ changes from 0.03 to 4. Following the pump pulse, the TPC signal quickly decays on a sub-ns time scale. This occurs due to fast trapping of photogenerated electrons by intragap states (upper inset). In these structures, photogenerated holes are immobile because of a large energy barrier created by the CdSe shell (lower inset). (c) Blue symbols show the ratio of the photocurrent measured at 1 ns after excitation ($B_{1ns}$) and the early time signal amplitude ($A$) for reference (noninverted) PbSe/CdSe QDs. Red symbols show a much higher ratio for inverted CdSe/PbSe QDs measured at 3 ns (see TPC traces in 'd'). (d) Pump-power dependent tail-normalized TPC traces for EDT-treated inverted CdSe/PbSe QDs (ZB CdSe cores with $r$ = 2 nm and shell

thickness of 0.9 nm; same sample as in Figure 2a) for ⟨$N$⟩ ranging from 0.04 to 6. In addition to the initial fast component due to electron trapping (upper inset), the TPC traces contain a long-lived tail indicating dot-to-dot charge transport due to mobile holes (lower inset). The increase in the relative amplitude of the initial fast TPC component at high pump fluences (⟨$N$⟩ > 1) is due to nonradiative Auger recombination.

To elucidate the feasibilities of applications of the developed inverted QDs in photoelectric devices, we investigate transient photocurrent (TPC) in QD solid-state films using Auston switch devices employing QDs as a photoactive layer.[51-53] These devices comprise a co-planar 50 Ω microstrip transmission line consisting of a gold layer (a ground plate) deposited onto the back side of a glass substrate and two top gold contacts separated by a 70 μm gap bridged by a film of QDs prepared *via* a layer-by-layer deposition approach (Figure 5a; see Experimental Section for device fabrication details). To facilitate dot-to-dot charge transport, the QDs were cross-linked using solid-state ligand exchange to replace original long capping molecules with short bi-functional molecules of 1,2-ethanedithiol (EDT). These procedures are similar to those previously applied to fabricate Auston switches based on PbSe QDs.[52,53]

For TPC measurements, the device was biased using a 40-to-100 V voltage applied between one of the top contacts and the ground plane. The QDs in the gap region were excited by 1.2 eV, 110 fs pulses from an amplified femtosecond laser (Experimental Section). A TPC launched by the laser pulse was time resolved with a 20-GHz oscilloscope used to monitor a transient voltage pulse generated on the 50 Ω resistor connected to the unbiased top contact and the ground plate (Figure 5a). The temporal resolution of this system is ~50 ps.

Previously, the Auston-switch technique has been used to study charge-carrier transport in QD films as well as carrier relaxation dynamics due to both geminate and nongeminate recombination.[53] Further, because of its high temporal resolution, it allows one to resolve multi-carrier decay due to Auger recombination, which was exploited for detecting and quantifying CM in device-grade QD films.[52]

First, we use the Auston-switch approach to investigate a TPC response of films made of traditional (noninverted) PbSe/CdSe QDs (see Experimental Section for fabrication details and Figure S6, Supporting Information for optical absorption and emission spectra). In Figure 5b, we display TPC traces measured for a range of pump fluences that correspond to excitation of 0.06 to 4 excitons per dot per pulse on average ($\langle N \rangle$). The recorded traces exhibit a pronounced increase in the amplitude with increasing pump fluence, which is accompanied by subtle changes in the dynamics. For all pump levels, the decay occurs on a short time scale of hundreds of picoseconds. Closer examination of photocurrent relaxation indicates that at low sub-single-exciton pump levels ($\langle N \rangle$ < 0.1), the signal half-life is ~170 ps and it shortens to ~70 ps at the highest pump level ($\langle N \rangle$ = 4). This shortening occurs due to the increasing fraction of the QDs populated with two or more excitons. In such dots, carrier decay becomes contributed by fast nonradiative Auger recombination during which the electron-hole recombination energy is transferred to a third carrier (an electron or a hole) residing in the same QD.[52,54]

The low-pump-intensity TPC dynamics ($\langle N \rangle$ < 0.1) observed for PbSe/CdSe QD samples are similar to the initial sub-ns photocurrent relaxation reported previously for plain, core-only PbSe QDs for which it was assigned to fast electron trapping by intragap states[53] (Figure 5b, upper inset) forming a weakly conducting mid-gap band. This band is responsible for 'dark' charge transport in non-illuminated films.[55] It also plays an important role in photoconductance as it controls relaxation of photoinjected mobile carriers.[53]

According to studies of ref. [53], in addition to the initial fast dynamics, core-only QD films exhibited a slow-relaxing component which was explained by nongeminate electron-hole recombination accompanying dot-to-dot migration of photoexcited holes. These mobile holes were identified as primary species responsible for photoconductance in PbSe QD films. The slow hole-transport-related dynamics are virtually absent in the films of reference (noninverted) PbSe/CdSe QDs (Figure 5b). In particular, the TPC signal measured at a time of $\Delta t = 1$ ns after the pump pulse ($B_{1ns}$) drops to ~ 10% of the early time signal amplitude ($A$) and becomes undetectable at $\Delta t = 3$ ns (Figure 5b and 5c, blue symbols). This is a direct consequence of core localization of photogenerated holes which inhibits their transport due to a high energetic barrier created by the CdSe shell (Figure 5b, lower inset).

The films of inverted CdSe/PbSe QDs exhibit a distinct behavior. In particular, in addition to fast (sub-ns) initial relaxation due to electron trapping (and Auger recombination at ($\langle N \rangle > 1$), they exhibit a pronounced slow-relaxing component (Figure 5d) in sharp contrast to the reference PbSe/CdSe QD sample. At low (sub-single-exciton) pump levels, the amplitude of the long-lived component measured at $\Delta t = 3$ ns ($B_{3ns}$) is approximately 70% of the peak photocurrent (Figure 5c, red symbols). This slow component is similar to that observed in previous studies of core-only PbSe QDs[53] where it was ascribed to charge transport dominated by mobile band-edge holes. This suggests that in films of the inverted CdSe/PbSe QDs, photogenerated holes are also mobile and, if necessary, can be accessed electrically (Figure 5d, lower inset). Thus, the developed inverted core/shell structures indeed exhibit the behavior we have aimed to achieve in the present work.

## 6. Conclusions and Outlook

To summarize, we have developed a novel WL based approach for uniform growth of a PbSe shell on top of CdSe cores. The synthesized core/shell QDs exhibit an inverted geometry for which a core of a wider gap material (CdSe) is enclosed into a shell of a narrower gap semiconductor (PbSe). Due to this arrangement of the two materials, both an electron and a hole exhibit a significant degree of shell localization, which makes them readily accessible electrically. The developed structures are well suited for applications that require extraction of carriers from the QD (*e.g.*, photochemistry) or carrier transport in the QD film (*e.g.*, photovoltaics and photodetection). To illustrate the latter capability, we incorporate our inverted QDs into photoconductive Auston switches and use these devices to demonstrate that the photoexcited current exhibits a long-lived component due to the slow nongeminate recombination of mobile photocarriers. This is in contrast to conventional noninverted PbSe/CdSe QDs, for which the detected photoresponse is short-lived, indicating that the photocarriers are immobile and quickly recombine in the geminate fashion in the initially excited QDs.

The developed QDs are of considerable interest for implementing CM-based advanced photoconversion schemes. In particular, standard noninverted PbSe/CdSe QDs have shown strong CM performance in all-optical spectroscopic measurements.[20] The newly developed inverted QD are ideally suited for revisiting the CM topic using electro-optical or photochemical studies. Such studies would help elucidate the relevance of the developed nanostructures to applications in photovoltaics and photochemistry.

Another interesting direction for future studies is exploration of a novel concept of spin-exchange CM (SE-CM).[50] Recently, high efficiency SE-CM was demonstrated using Mn-doped PbSe/CdSe QDs.[50] It was proposed that SE-CM occurred via two steps: (1) fast SE-mediated excitation transfer from a CdSe shell to a Mn ion, followed by (2) its spin-conserving relaxation back to the ground state by creating two excitons ('dark' and 'bright') in the PbSe core. This process was detected using femtosecond transient absorption and time-resolved PL which were also applied to quantify CM yields. Upon their doping with Mn, the novel inverted QDs can, in principle, be used to advance the work on SE-CM from purely fundamental optical studies to electro-optical investigations of practical relevance to photoelectric devices.

We expect that this WLAD synthesis protocol can be extended to create a wide range of other heterostructured QDs that are difficult to realize using existing synthetic methods. For example, traditional noninverted PbSe/CdSe QDs can be coated with a PbSe(S) outer shell to create structures with type (I+II) characteristics useful for applications as optical gain media.[56] This will facilitate the development of NIR QD lasers and optical amplifiers applicable, for example, in telecommunications and silicon photonics. In addition, the developed WLAD approach can simplify the integration of dissimilar materials (*e.g.,* semiconductors and metals) into a single nanocrystal, leading to mixed-functionality 'hybrid' structures with interesting and potentially useful properties.

## 7. Experimental Section

*Materials*: Lead (II) oxide (PbO, 99.999%, Aldrich), lead acetate trihydrate (≥99.99%, Aldrich), oleic acid (OA, 90%, Aldrich), trioctylphosphine (TOP, 97%, STREM), oleylamine (OLA, 70%, Aldrich), selenium powder (Se, 200 mesh, 99.999%, Alfa Aesar), trioctylamine (98%, Aldrich), cadmium oxide (CdO, 99.99%, Aldrich), cadmium chloride ($CdCl_2$, 99.99%, ACROS Organics), sodium selenide ($Na_2Se$, 99.8%, Alfa Aesar), *N*-octadecylphosphonic acid (ODPA, >97%, Aldrich), cadmium nitrate tetrahydrate (99.999%, Aldrich), myristic acid (98.5%, Aldrich), selenium dioxide ($SeO_2$, 99.999%, Aldrich), 1,2-hexadecandiol (90%, Aldrich), octadecene (ODE, 90%, Aldrich), *N*-methylformamide (MFA, 99%, Aldrich), *N*-butanol (99.8%, Aldrich) and other solvents and non-solvents were used as received. Octadecene and MFA were dried at 80° C for 10 hours under vacuum and stored in a glovebox.

*Preparation of CdSe cores*: Monodisperse CdSe QDs with the WZ crystal structure in the size range of 2–4 nm were synthesized following ref. [25]. Synthesis of ZB CdSe QDs in a size range of 3–4 nm was performed *via* a previously described method of ref. [26]. After the synthesis, the QDs were precipitated with either acetone or *N*-butanol (one time) and ethanol (twice) and redispersed in hexane with a concentration of ~150 mg mL$^{-1}$. The concentration was calculated based on the measured absorption coefficient and QD extinction coefficients from ref. [57].

*Preparation of a wetting layer*: All synthetic steps were performed in a glove-box under oxygen-free and moisture-free conditions using c-ALD at room temperature. Original organic ligands of the CdSe QDs (20–50 mg in 3 mL of hexane) were replaced with Se$^{2-}$ ionic species (0.1 M solution of Na$_2$Se in 3 mL of MFA) using a solution-phase ligand exchange reaction leading to the formation of a stable colloid of the Se$^{2-}$-capped CdSe QDs. The Se$^{2-}$-capped CdSe QDs were precipitated twice using MFA–acetone/toluene as a solvent–non-solvent pair to remove the unreacted precursors and remaining reaction byproducts. Afterward, the QDs were transferred to polar MFA for growth of the Pb half-layer of a WL. 0.1–0.2 mL of 0.5 M Pb$^{2+}$ cation precursor solution (in the form of lead-acetate in MFA) was added to diluted Se$^{2-}$-capped CdSe QDs (3–6 mg in 5 mL of MFA), and the mixture was stirred for 2–3 min. It is important to use diluted Se$^{2-}$-capped CdSe QDs to prevent flocculation or gelation of the QDs due to the collapse of the diffuse counterion cloud around the electrostatically stabilized QDs. The CdSe/Se$^{2-}$/Pb$^{2+}$ QDs consisting of a CdSe cores and a thin PbSe layer were precipitated one time using MFA–acetone/toluene as a solvent–non-solvent pair. These procedures produced a stable solution of the CdSe/Se$^{2-}$/Pb$^{2+}$ QDs in MFA with a concentration of ~60 mg mL$^{-1}$.

*Transfer of* CdSe/Se$^{2-}$/Pb$^{2+}$ QDs *into a non-polar solvent*: The next step was the surface functionalization of CdSe/Se$^{2-}$/Pb$^{2+}$ QDs with oleate ions (introduced as lead-oleate) to assist the transfer of the QDs from polar MFA into a non-polar medium. The oleate solution was prepared by dissolving 953.6 mg of PbO and 3.5 mL of oleic acid in 20 mL of ODE, degassed by heating to 110 °C under vacuum and then, under nitrogen at 150 °C for 1 and 2 hours, respectively. 3 mL of the stock oleate solution in ODE was added to dry CdSe/Se$^{2-}$/Pb$^{2+}$ QDs, which were precipitated from MFA. After mild ultrasonification, the QDs were transferred to ODE, leading to the formation of oleate-capped CdSe/Se$^{2-}$/Pb$^{2+}$ QDs. These structures are referred to in this work as functionalized (CdSe+WL) cores. The QDs were flocculated with ethanol and redispersed in ODE to form a stable concentrated solution (30–60 mg mL$^{-1}$).

*High-temperature PbSe shell growth*: To grow a PbSe shell, lead and selenium precursors were prepared in advance. The lead-precursor solution was prepared by dissolving 444.2 mg of PbO in 2 mL of oleic acid and 10 mL of ODE at 150 °C under nitrogen flow (for 2 hours) and subsequently dried under vacuum at 110 °C (for 1 hour). The selenium-precursor solution was prepared by dissolving 1.5 mL of TOP·Se (1 M) in 4 mL of TOP.

To prepare a PbSe shell with up to a 3-ML-thickness on top of ZB CdSe cores with a radius of 1.9 nm, 700 μL of (CdSe+WL) seeds with a concentration of 5·10$^{-4}$ M in ODE in one syringe and a solution of 1.5 mL of 1 M TOP·Se in 4 mL of TOP in another syringe were simultaneously injected into the dried solution of 9 mL of lead-precursor, 16 mL of ODE and 4 mL of oleic acid at 130 °C. The temperature of the reaction mixture was increased up to 150 °C within 3 mins followed by fast cooling with a water bath.

To grow a PbSe shell up to a 4-ML thickness on top of WZ CdSe cores with a radius of 1.9 nm, 1 mL of freshly prepared (CdSe+WL) seeds with a concentration of 5·10$^{-4}$ M in ODE in one syringe and a solution of 1.5 mL of 1 M TOP·Se in 4 mL of TOP in a separate syringe were injected into the dried solution of 9 mL of lead-precursor, 16 mL of ODE, and 4 mL of oleic acid at 130 °C. The temperature of the reaction mixture was increased up to 160 °C within 4 mins followed by fast cooling with a water bath.

To grow thicker shells ($H > 3$ ML for ZB cores, and $H > 4$ ML for WZ cores), multiple injections of small amounts of lead- and selenium-precursors were performed at 140 °C within 10 min. The synthesized core/shell QDs were precipitated from solution by adding a minimum amount of ethanol-acetone (1/3 V/V). The QDs were redispersed in toluene and stored in a nitrogen glovebox.

*Synthesis of noninverted core/shell PbSe/CdSe QDs*: Preparation of PbSe QDs. 476.8 mg of PbO and 2.74 mL of oleic acid were dissolved in 15 mL of ODE. The mixture was heated to 150 °C for 0.5 h under nitrogen to form a clear solution of lead (II) oleate and degassed at 100 °C under vacuum ($\sim 10^{-2}$ bar) for 1 h. Then, the resulting solution was heated to 170 °C. and a mixture of 6.75 mL of 1 M TOP·Se and 35 µL of diphenylphosphine was quickly injected into it. The PbSe QDs were grown for 2 min followed by an abrupt cooling using ice-water. To precipitate the synthesized particles, ethanol was added to the solution. The colloid was further centrifuged, and the precipitate was redispersed in pure hexane. This cleaning step was repeated twice. Finally, the PbSe QDs capped with oleic acid were redispersed in octane and stored in a glovebox.

Cation-exchnage reaction to prepare a CdSe shell. Core/shell PbSe/CdSe QDs were preapred using cation exchange of $Pb^{2+}$ for $Cd^{2+}$ withint the surface layer of the synthesized PbSe QDs. To prepare Cd-oleate, a mixture of 0.64 g of CdO, 5 mL of oleic acid, and 5 mL of ODE was heated to 260 °C and kept at this temperature for 1 hr. The resulting solution was placed under vacuum at 110 °C for 1 hour. PbSe QDs in hexane were slowly added into the cadmium-oleate mixture at 60 °C, and then hexane was completely removed under vacuum. A cation exchange reaction was conducted at 120 °C for 16 hrs to form PbSe/CdSe QDs. The washing of PbSe/CdSe QDs was performed twice using ethanol-toluene as a solvent/non-solvent pair.

*Structural and compositional studies of QDs:* Transmission electron microscope (TEM) and high-resolution TEM (HR-TEM) images were taken using a JEOL 2010 TEM. QD samples were prepared as drop-cast films deposited onto carbon-coated copper grids.

Powder X-ray diffraction (XRD) patterns were collected with a PANalytical EMPRYREAN diffraction system in the reflection mode. A nickel filter, Cu K$\alpha$1 irradiation, and a PIXcel 1D detector were used in the measurements.

The chemical composition of the QDs was determined using a Shimadzu ICPE-9000 ICP-OES system.

*Sample preparation for spectroscopic measurements*: During sample preparation, QDs were handled in a nitrogen glove box to avoid oxidation. The purified QDs dispersed in trichloroethylene (TCE) were loaded into a 1 mm thick air-tight quartz cuvette and tightly sealed. QD samples were diluted such that to obtain an optical density of ~0.05 at the position of the band-edge transition. During time-resolved PL measurements, the QD solution was continuously stirred to avoid uncontrolled photocharging and sample degradation.

*Optical absorption and steady state PL measurements*: Absorption spectra were measured by a PerkinElmer Lambda 950 spectrophotometer. Visible PL spectra were recorded using a Horiba Scientific FluoroMax-4 spectrofluorometer with a 400 nm excitation source. NIR PL spectra were taken using a home-built system with a 532 nm laser as a source of excitation. QD emission was analyzed by scanning a grating monochromator coupled to a liquid nitrogen cooled InSb detector. A NIR PL QY was obtained using IR-26 dye (QY = 0.048%) as a reference.

*Time-resolved PL measurements*: The QD samples were excited using ~100 fs pulses of either the fundamental output (1.55 eV photon energy; NIR PL measurements) or the second harmonic output (3 eV photon energy; visible PL measurements) of a femtosecond Ti:sapphire amplified laser (Coherent RegA 9000). The QD emission was spectrally filtered using a monochromator and detected by a superconducting single-nanowire single-photon detector (SNSPD) cooled to 3.5 K. A width of the instrument response function of this system was ~60 ps.

*Fabrication of Auston switch devices*: 100 nm of gold was evaporated onto the back of a glass substrate. A 200-nm thick QD film (inverted CdSe/PbSe or noninverted PbSe/CdSe QDs) was deposited on top of the substrate *via* repeated spin-coating/ligand exchange/rinsing steps. During the deposition procedure, the QDs dispersed in octane (20 mg mL$^{-1}$) were spin-coated at 1600 rpm for 30 s. For ligand exchange, 60 µL of a 1 M EDT solution in MeOH was dropped onto the QD layer, soaked for 60 s and then, spin-coated at 1200 rpm for 30 s. For rinsing, pure MeOH was spin-coated onto the QD layer to remove residual excess ligands. To prepare a 200-nm thick QD layer, these three steps were repeated 5 times. The resulting QD film was annealed for 30 min at 150 °C. Two gold contacts (200 nm-thickness) separated by a 70 µm gap were evaporated on top of the QD film through a shadow mask. Together with the gold plate at the bottom of the substrate, these contacts form a co-planar microstrip transmission line with nominal impedance of 50 Ω.

*Transient photocurrent measurement*: The as-prepared devices were loaded into a nitrogen-filled air-tight housing to avoid QD oxidation. One of the top gold contacts was connected using coaxial cables to a sampling oscilloscope with a 20 GHz bandwidth, while the other contact was biased using a direct-current voltage source. QDs in the gap between the gold contacts were excited by 110-fs, 1.2-eV pulses of a regeneratively amplified ytterbium-gadolinium tungstate (Yb:KGW) femtosecond laser (Pharos, Light Conversion). The pulse repetition rate was 20 kHz. The laser beam was focused with a cylindrical achromatic lens into a stripe with dimensions of ~2 mm by ~300 µm, which were much larger than the device photoactive area (500 µm by 70 µm) and, as a result, allowed us to achieve spatially uniform excitation. To minimize timing jitter, the oscilloscope was triggered by laser pulses using a high-speed Si photodetector.

## Supporting Information

Supporting Information is available from the Wiley Online Library or from the authors.

## Acknowledgements

These studies were supported by the Solar Photochemistry Program of the Chemical Sciences, Biosciences and Geosciences Division, Office of Basic Energy Sciences, Office of Science, U.S. Department of Energy.

## Conflict of Interest

The authors declare no conflict of interest

## Data Availability Statement

The data that support the findings of this study are available from the corresponding author upon reasonable request.

## Keywords

# Supporting Information for

**Wetting-Layer-Assisted Synthesis of Inverted CdSe/PbSe Quantum Dots and their Photophysical and Photo-Electrical Properties**

*Vladimir Sayevich, Whi Dong Kim, Zachary L. Robinson, Oleg V. Kozlov, Clément Livache, Namyoung Ahn, Heeyoung Jung, and Victor I. Klimov[*]*

Nanotechnology and Advanced Spectroscopy Team, C-PCS, Chemistry Division, Los Alamos National Laboratory, Los Alamos, New Mexico 87545, United States

E-mail: klimov@lanl.gov

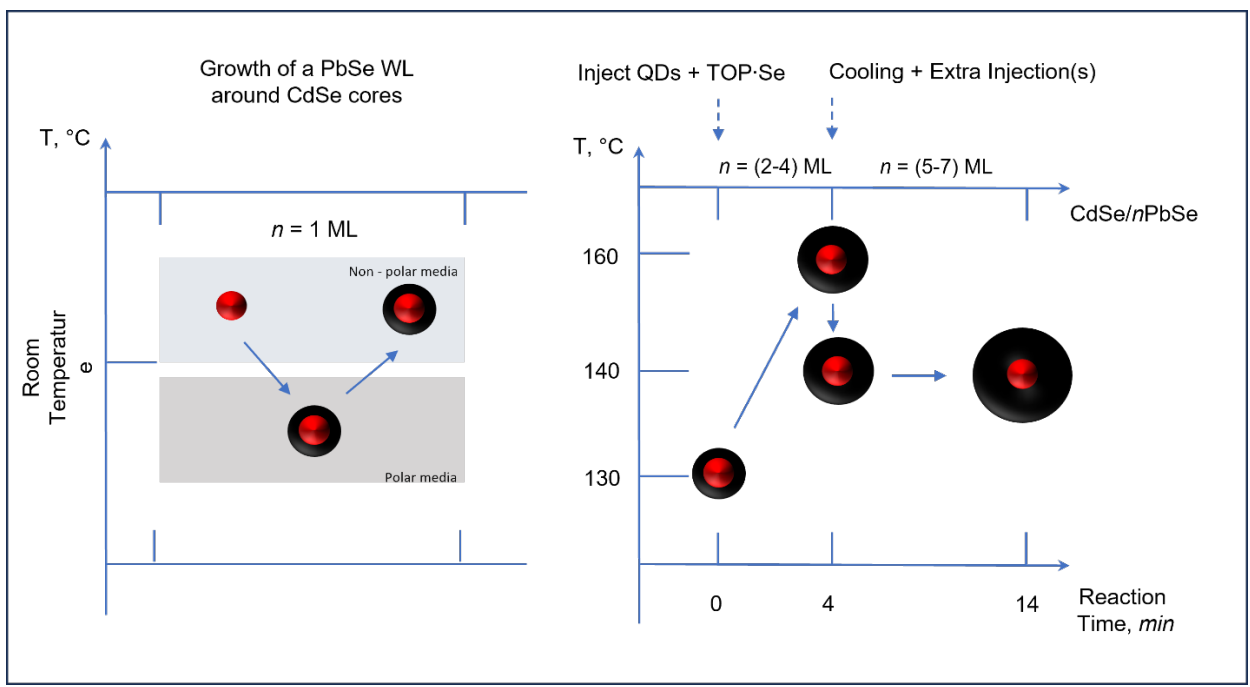

Scheme S1. Schematic illustration of the shell growth process. Starting with CdSe cores at room temperature, original organic ligands are replaced with $Se^{2-}$ ionic species during quantum dot (QD) transfer from a non-polar medium (ODE) to a polar solvent (MFA) using solution-phase ligand exchange. Then colloidal atomic layer deposition (c-ALD) is applied to deposit a layer of $Pb^{2+}$ cations, which completes a Se-Pb wetting layer (WL). After addition of lead-oleate, the $CdSe/Se^{2-}/Pb^{2+}$ QDs are transferred back to a non-polar medium (ODE) to form oleate-capped $CdSe/Se^{2-}/Pb^{2+}$ QDs referred here to as 'functionalized (CdSe+WL) cores.' To grow the PbSe shell, the (CdSe+WL) particles and Pb and Se precursors are injected into a mixture of ODE and oleic acid at 130°C. The temperature of the reaction is increased to 160°C within several minutes which leads to shell thicknesses from 2 to 4 PbSe monolayers (MLs). The shell thickness can be further increased to 5–7 MLs by fast cooling to 140°C followed by extra injections of Pb and Se precursors.

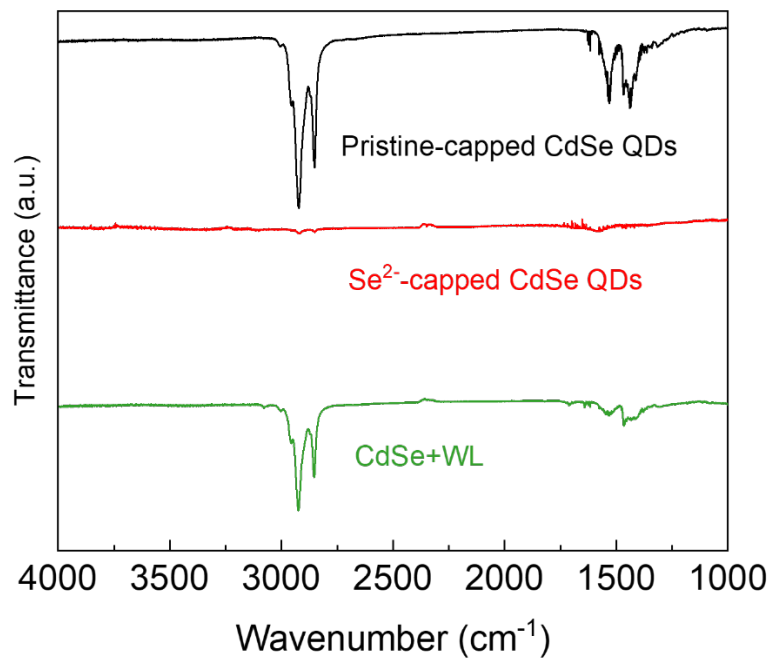

Figure S1. Fourier transform infrared (FTIR) spectra of WZ core-only CdSe QDs, with a radius of $r$ = 1.9 nm capped with original organic ligands (black trace) and after ligand exchange for inorganic $Se^{2-}$ species and transfer into a polar phase (red trace). Green trace is the FTIR spectrum of QDs enclosed into a PbSe WL (CdSe+WL) and transferred back into a nonpolar phase with oleate co-ligands. The spectra are offset vertically for clarity. The characteristic C-H stretching modes (2700–3000 cm$^{-1}$) completely disappear following ligand exchange with $Se^{2-}$ species, confirming a complete removal of original organic capping groups. The C-H stretching modes re-appear after the (CdSe+WL) particles are transferred back into a nonpolar solvent.

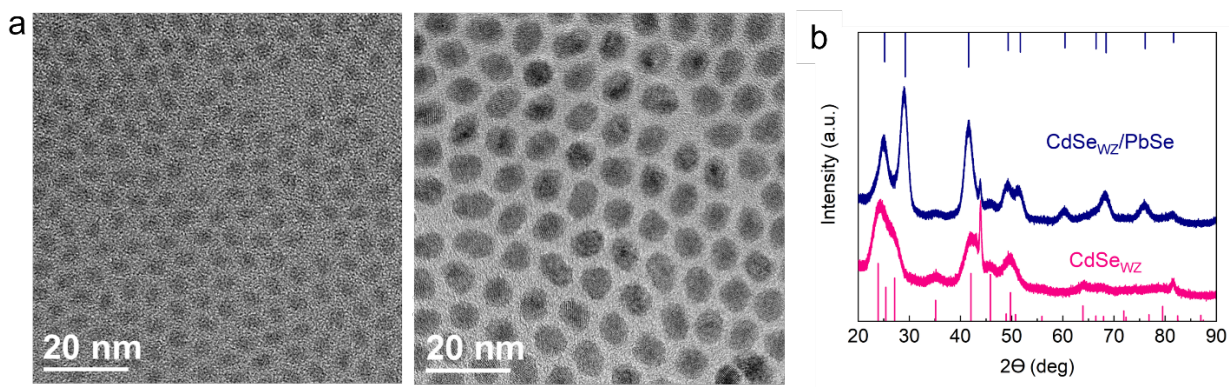

Figure S2. (a) Transmission electron microscopy (TEM) images of starting wurtzite (WZ) CdSe cores with a radius of $r$ = 1.9 nm (left) and final CdSe/PbSe core/shell QDs (right) with a PbSe shell thickness ($H$) of approximately 2 nm or ~7 semiconductor MLs. The overall radius of the core/shell QDs ($R$) is approximately 3.5 nm. (b) The corresponding X-ray diffraction patterns of the core (pink) and the core/shell (blue) samples. Vertical bars show positions of diffraction peaks of bulk WZ CdSe (bottom; pink) and bulk rock salt PbSe (top; blue).

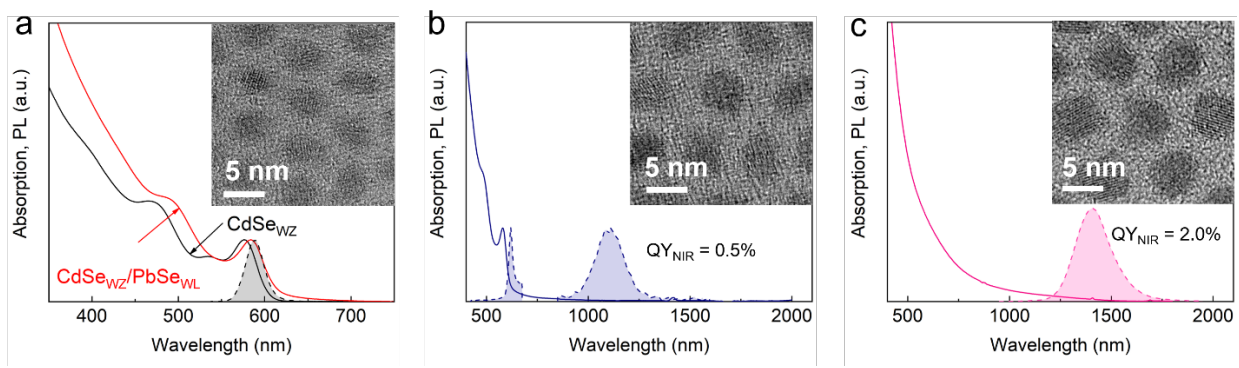

Figure S3. Linear absorption (lines), steady-state photoluminescence (PL, shaded spectra), and corresponding high-resolution TEM images (insets) of (a) WZ CdSe cores (black lines) with a radius of $r$ = 1.9 ± 0.2 nm, (b) CdSe/PbSe QDs for a shell-growth time of $t_{sh}$ = 0.5 min ($H$ = 0.4 nm or ~1.3 ML), and (c) $t_{sh}$ = 4 min ($H$ = 1.2 nm or ~4 MLs). $QY_{NIR}$ is the quantum yield of the near-infrared (NIR) PL band. The red line in (a) is the absorption spectrum of non-emissive CdSe/PbSe QDs which contain a 1 ML thick PbSe wetting layer (WL).

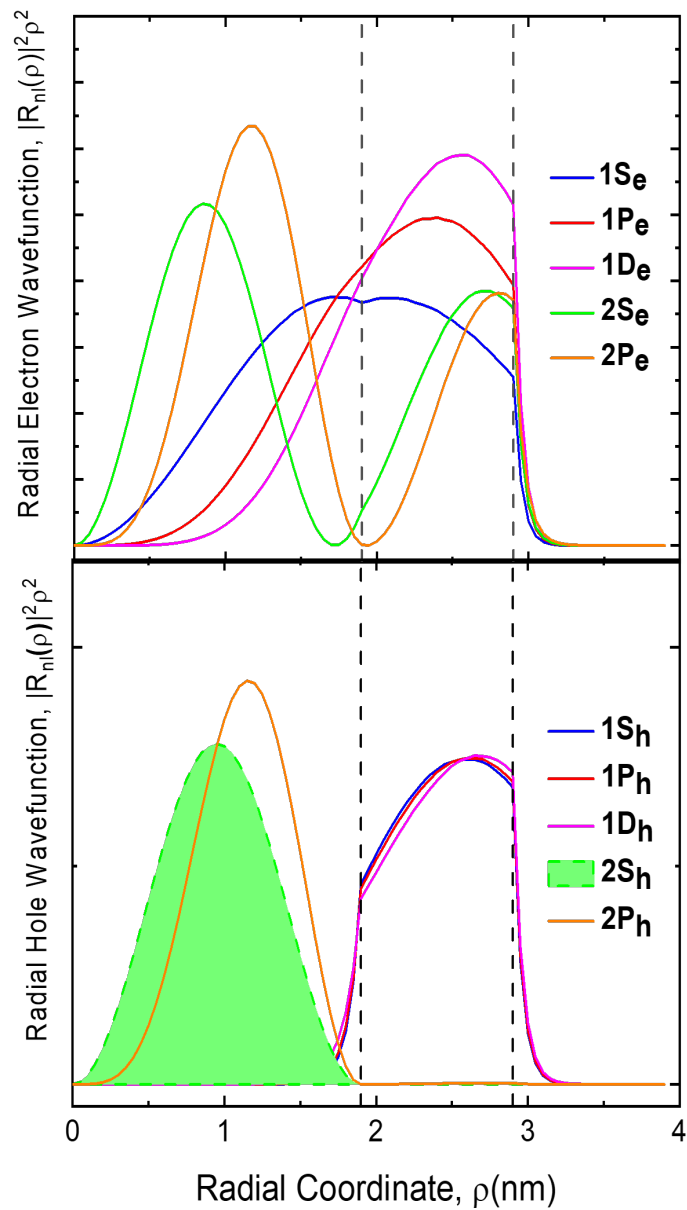

Figure S4. Calculated radial electron and hole wavefunctions ($R_{nl}(\rho)$) for five lower-energy electron and hole states ($1S_{e,h}$, $1P_{e,h}$, $1D_{e,h}$, $2S_{e,h}$, and $2P_{e,h}$) in inverted CdSe/PbSe QDs with $r$ = 1.9 nm and $H$ = 1 nm.

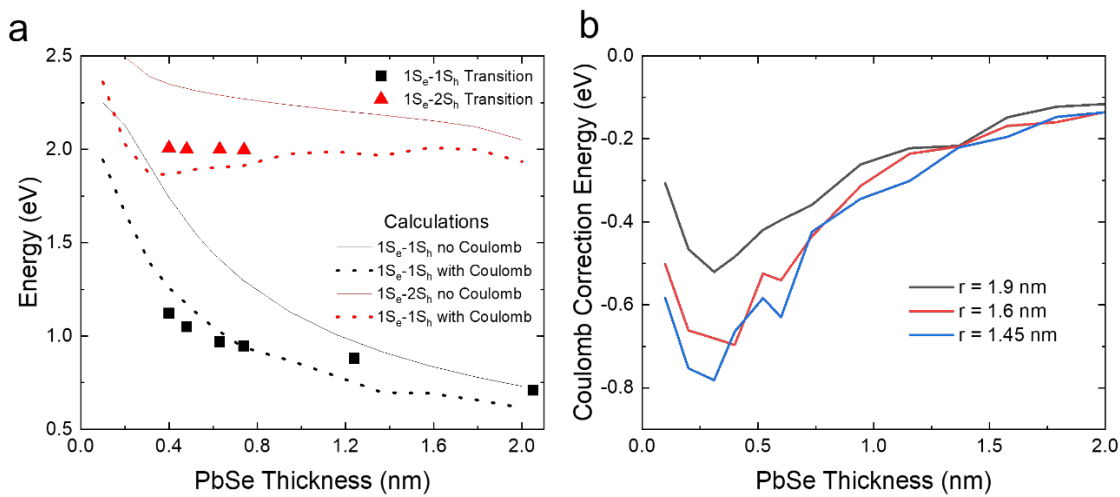

Figure S5. (a) Calculated $1S_e$–$1S_h$ (black) and $1S_e$–$2S_h$ (red) transition energies as a function of shell thickness for $r$ = 1.9 nm without accounting for Coulomb interactions (solid lines) and with the electron-hole Coulomb interaction included (dashed lines) as is shown in the main text. Symbols are the spectral energies of the visible (red triangles) and NIR (black squares) PL bands from Figure 3. (b) The electron-hole Coulomb interaction energy for three different core radii ($r$ = 1.45, 1.6, and 1.9 nm). The oscillations come from imposing a finite grid on sub-ML spacings. Decrease in the magnitude of the Coulomb correction for small PbSe shell thicknesses (<0.3 nm) is due to weak confinement of holes in a sub-ML PbSe layer.

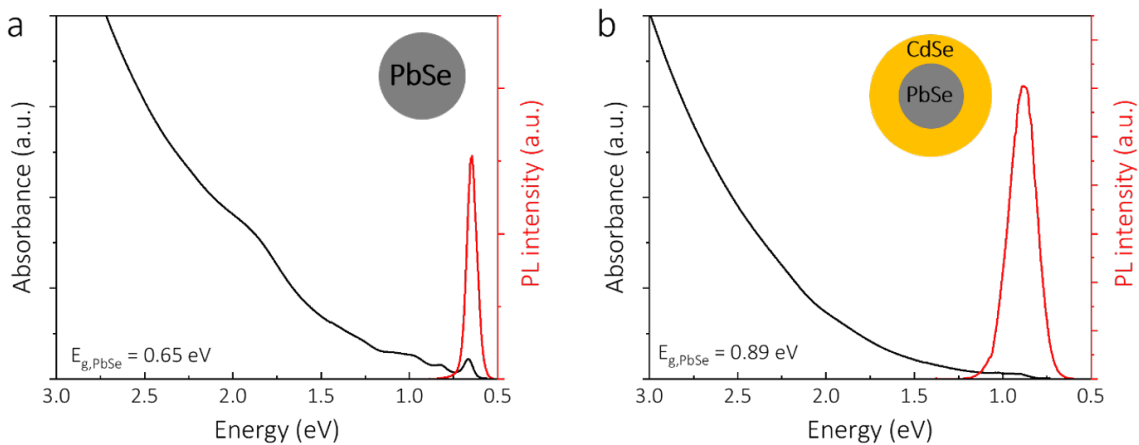

Figure S6. PL (red line) and absorption (black line) spectra of PbSe cores (a) and noninverted PbSe/CdSe core/shell QDs (b). The overall QD radius is ~4 nm and the shell thickness is ~2.5 nm. These core/shell QDs were used as the reference sample in TPC studies shown in Figure 5.